# Mega-Gauss Plasma Jet Creation Using a Ring of Laser Beams

L. Gao[1], E. Liang[2], Y. Lu[2], R.K. Follet[3], H. Sio[4], P. Tzeferacos[5], D.H. Froula[3], A. Birkel[4], C. Li[4], D. Lamb[5], R. Petrasso[4], W. Fu[2], M. Wei[6], H. Ji[1]

[1]Princeton Plasma Physics Laboratory, Princeton, NJ 08536

[2]Rice University, Houston, TX 77005

[3]Laboratory for Laser Energetics, Rochester, NY 14623

[4]Massachusetts Institute of Technology, Cambridge, MA 02139

[5]University of Chicago, Chicago, IL 60637

[6]General Atomics, San Diego, CA 92121

## Abstract

Using 20 OMEGA laser beams at the Laboratory for Laser Energetics, University of Rochester, to irradiate a flat plastic target in a hollow ring configuration, we created supersonic cylindrical stable plasma jets with self-generated megagauss magnetic fields extending out to > 4 mm. These well-collimated magnetized jets possess a number of distinct and novel properties that will allow us to study the dynamics, physical processes and scaling properties of astrophysical jets not feasible with other laboratory settings. The dimensionless parameters of these laboratory jets fall in the same regime as those of YSO jets. They will also provide new versatile laser-based platforms to study magnetized shocks, shear flows and other plasma processes under controllable conditions.

## 1. INTRODUCTION

Magnetized plasma jets are ubiquitous in the universe (Ferrari 1998, Livio 1999, Sari et al 1999, Bally et al 2007, Frank et la 2014). It is thus highly desirable to recreate them in the laboratory to study their physical processes and scaling properties under controllable conditions. Recently, irradiation of solid targets with high-energy lasers has become a popular tool to launch supersonic plasma outflows for a broad range of applications, including the study of astrophysical jets (see articles in Hartigan 2013, Ciardi 2015).

When an intense laser irradiates a solid target, strong toroidal magnetic fields are created around the laser spot by the $\nabla P_e \times \nabla n_e$ ("Biermann battery") term (Biermann 1950) of the generalized Ohm's law (Krall and Trivelpiece 1973). However, these magnetic fields are localized at the surface, spanning distances ~ only a few times laser spot size. They decay rapidly in time and space as the outflow expands, diverges and rarifies. Hence the creation of stable plasma jets with strong self-generated magnetic fields at large distances from the laser target remains an unsolved challenge. Here we report a new laser platform capable of creating stable plasma jets with megagauss (MG) self-generated fields extending to > 4 mm, by using 20 OMEGA laser beams at the Laboratory for Laser Energetics (LLE, Boehly et al 1995) of the University of Rochester, to irradiate a flat CH target in a hollow ring pattern.

While large-scale magnetized jets have been created using pulse power machines by many groups (Lebedev, et al. 2002, 2005, Suzuki-Vidal et al. 2011, Gourdain et al. 2010, Albertazzi, et al. 2014), these jets consist of low-density plasmas ($n_e \leq 10^{19}/cm^3$) accelerated by externally induced **j X B** forces and collimated by strong toroidal fields. Hence their dynamics and properties are fundamentally different from those of hydrodynamic jets launched by laser-solid-target interactions. Laser-driven hydrodynamic jets have also been created recently, using V-shaped foil targets to thermalize the transverse momentum and facilitate axial collimation (Gregory et al. 2008, Li et al 2016). However, these jets are created with inherently non-cylindrical configurations and are often unstable (Li et al 2016). Hence the dynamics and stability of cylindrical jets produced in most common astrophysical settings cannot be faithfully studied using the V-wedge platform. In contrary, collimated jets created with a ring of laser beams irradiating flat targets reported in this paper are inherently cylindrical and stable to first order. As we discuss below, ring-laser jets have many distinct properties not achievable using pulse power or V-wedge laser targets. Hence the ring-laser-jet platform presented here is complementary to these other jet-launching mechanisms.

Our OMEGA experiment was originally motivated by 2D cylindrical FLASH (https://flash.uchicago.edu) simulations, which showed that a hollow-ring-laser jet can reach much higher density and temperatures along the jet axis due to compression and heating of the convergent on-axis flow (Fu et al 2013), compared to outflows launched by lasers irradiating a single spot on the target. Furthermore, a hollow-ring-laser jet was shown to create and sustain magnetic fields at large distances from the target (Fu et al 2015). When the ring-shaped laser-driven blow off collides along the axis, strong gradients in electron density $n_e$ and electron pressure $P_e$ produce “Biermann battery” (Biermann 1950) magnetic fields along the axis. The larger the hollow ring radius, the stronger the field becomes and the further it extends from the target (Fu et al 2015). To demonstrate this hollow-ring-laser jet concept, we carried out a series of experiments in 2015 and 2016 at the OMEGA laser facility of LLE (Boehly et al 1995). 20 beams from the upper hemisphere of the OMEGA facility were arranged to form a ring pattern at the flat CH target at laser intensities > $10^{14}$ W/cm$^2$ (Fig.1), with ring radius d ranging from 0 μm to 1200 μm. We used proton radiography (P-rad) (Li et al 2006, Zylstra et al 2012) to diagnose the magnetic field, optical Thomson scattering (TS) (Froula et al 2006, Katz et al 2012, Mackinnon et al 2004, Follett et al 2016) to measure the plasma parameters at the target chamber center (TCC, 2.5 mm above target, Fig.1), and time-lapse imaging with an x-ray framing camera (XRFC, Bradley et al 1995, Benedetti et al 2012) to image the jet emission. The experimental setup is sketched in Fig.1.

## 2. RESULTS ON MAGNETIC FIELDS

The most interesting and important results come from the diagnostic of magnetic fields. Fig.2 shows raw P-rad images from laser rings of radius d=0, 400, 800 and 1200 μm. As the radius d increases, the strongly magnetized plasma extends to ~5 mm for d = 1200 μm. The light and dark patterns, created by proton deflections, correspond to net positive and negative currents projected along the line of sight into and out of the plane (Kugland et al 2012, Graziani et al 2016, Bott et al 2017). Hence vertical P-rad filaments correspond to $\mathbf{B}_z$ fields, while Y-shaped branches are mainly caused by $\mathbf{B}_\phi$ fields viewed at a slant angle. As the ring radius d is increased, stronger fields are created further from the target and more concentrated along the axis. Even though the true 3D magnetic field geometry cannot be uniquely inferred from the 2D proton images, we can still constrain the projected **B** fields integrated along the line of sight using direct inversion techniques (Graziani et al 2016, Bott et al 2017). Figure 3 gives a sample direct inversion result for the **B**-field of the d=800μm jet at 3.6ns along the blue line-out. Since the jet physical width deduced from x-ray images is < 1.1 mm, we conclude that the maximum orthogonal **B** field (central spike in Fig.3(c)) must exceed a MG.

Below we present the simulated P-rad images of the d=800 μm jet using **B**-fields predicted by 3D FLASH simulations. Details of the simulations will be reported elsewhere (Lu et al 2019). The generalized Ohm's law (Krall and Trivelpiece 1973) used in the FLASH code includes advection, diffusion and Biermann battery (Biermann 1950) terms (c=light speed, e=electron charge):

$$\frac{\partial \mathbf{B}}{\partial t} = \nabla \times (u \times \mathbf{B}) - c\nabla \times (\eta \mathbf{j}) + c\frac{\nabla P_e \times \nabla n_e}{e n_e^2} \quad (1)$$

where u is flow advection velocity, $\eta$ is electrical resistivity, $P_e$ is electron pressure and $n_e$ is electron density. Fig.4a shows |**B**| profiles at 3 ns predicted by FLASH for four different ring radii d. As d is increased, the maximum field increases and becomes more parallel and concentrated towards the axis, consistent with the P-rad images of Fig.2. The maximum FLASH-predicted fields reach ~ MG for d =800 μm and 1200 μm, also consistent with the direct inversion results (Fig.3). Detailed field line plots of the d = 800μm jet (Fig.4b) show that they are dominated by $\mathbf{B_z}$ fields near the jet axis and by $\mathbf{B_\phi}$ fields near the target. Fig.4c compares the FLASH-simulated P-rad images (left column) with the observed images (right column) at different times, showing good agreement. Both the spacing and contrast of the bright and dark streaks, which are sensitive to absolute field amplitudes, are consistent between FLASH simulation and experiment.

Detailed analysis suggests that the primary cause of seed field generation is the collisions between blow offs from individual laser spots, which create non-parallel density and temperature gradients on the scale of the laser spot radius r~125μm. These filamentary seed fields are then advected toward the jet axis and compressed, producing the strongest fields near the axis (Figs.4ab). When the ring radius d increases, there is more room for radial advection and compression, leading to stronger **B** fields (Fig.4a). At late times, the fields are dominated by a few para-axial bundles with $\mathbf{B_z} > (\mathbf{B_r}, \mathbf{B_\phi})$ (Fig.4b), which result in the proton images we see in Fig.2 & Fig.4c. Thus we have demonstrated the creation of cylindrical plasma jets with MG self-generated poloidal fields extending to > 4mm along the jet axis. The amplitude and geometry of these fields can be manipulated by dialing the ring radius and laser parameters.

We can obtain order-of-magnitude estimates of the maximum B-field near the jet axis from dimensional analysis using Eq.(1), by balancing the advection term with the Biermann battery term, since the diffusion term is negligible in this case. We find $B_{max} \sim (cd/eu)(kT_e/r^2)$ where u = radial advection velocity ~ ion thermal velocity~$(kT_i/Am_p)^{1/2}$, k=Boltzmann constant, $m_p$=proton mass, and A=6.5 for CH. Hence $B_{max} \sim$ MG (d/800μm) ($T_e$/keV)($T_i$/A/keV)$^{-1/2}$, in good agreement with both direct inversion results (Fig.3) and 3D FLASH predictions (Fig.4a). This scaling formula suggests that $B_{max}$ can be increased by increasing d or $T_e$ (which increases with laser intensity). We also note that in our previous 2D simulations (Fu et al 2015), because of the assumption of perfect cylindrical symmetry, the only gradient length scale is d. Replacing $d/r^2$ with 1/d in the equation above, we obtain $B_{max} \sim$ 25kG for d=800μm, again in good agreement with 2D FLASH predictions (Fu et al 2015). Besides the difference in the magnitude of $B_{max}$, a 2D-cylindrical Biermann battery can only generate $\mathbf{B}_\phi$ (Fu et al 2015), whereas the 3D Biermann battery field is complex and dominated by $\mathbf{B}_z$ near the axis (Fig.4b). We emphasize that simulated P-rad images using 25kG pure $\mathbf{B}_\phi$ fields completely disagree with the observed images (Fig.2 & Fig.4c).

## 3. PLASMA PARAMETERS AND JET MORPHOLOGY

The time histories of density, temperatures and velocity at TCC were measured using optical Thomson scattering (TS, Froula et al 2006, Katz et al 2012, Mackinnon et al 2004) with a $2\omega_o$ ($\lambda_o$ = 526.5 nm) probe beam. Fig.5 compares the experimental data derived from TS with 3D FLASH predictions. The plasma parameters were inferred from the TS spectra using the technique discussed in Follett et al (2016). Comparing with FLASH predictions, the agreements for electron density and flow velocity are excellent, and the qualitative trends for $T_i$ and $T_e$ are basically consistent. However, we still need to improve the temperature calculations in FLASH to get better quantitative agreement with the temperatures inferred from TS. Both simulation and experimental data suggest that the d=800 μm jet achieves the highest maximum axial density, whereas the d=1200 μm jet achieves the highest maximum axial temperature, respectively.

Evolution of the global jet morphology was observed using time-lapse X-ray imaging with an XRFC (Bradley et al 1995, Benedetti et al 2012) located at $38^o$ from the jet axis. Fig.6 compares the evolution of two d=800 μm jets, one with 2% Fe-doped CH target (Fig.6a) and one with pure CH target (Fig.6b). Both jets are well collimated and stable, but the Fe-doped jet appears narrower than the pure CH jet due to stronger radiative cooling, consistent with FLASH predictions. In principle, these x-ray images can be used to constrain the density and temperature profiles if the x-ray intensities were absolutely calibrated. Unfortunately, absolute calibration of the x-ray cameras was not performed for these experiments due to the shortage of shots. We plan to do it for future experiments.

## 4. DISCUSSIONS AND ASTROPHYSICAL APPLICATIONS

Our method of creating strongly magnetized cylindrical jets using a ring laser is ideal for scaling up by using more lasers and higher-intensity beams, such as those available at the National Ignition Facility (NIF) at Livermore CA. In terms of magnetic fields, the key advantage of using more laser beams (e.g. up to 64 beams at NIF) is to make the ring pattern larger and more uniform. A larger laser ring creates stronger fields due to more radial compression, and a more uniform ring produces larger pitch angle $\mathbf{B}_\phi/\mathbf{B}_z$, because better azimuthal symmetry enhances $\mathbf{B}_\phi$ and reduces $\mathbf{B}_z$. The magnetic pitch angle plays important roles in the stability and lateral expansion of the jet (Krall & Trivelpiece 1973). Besides modeling astrophysical jets, supersonic outflows with well-characterized magnetic fields can be used to study magnetized shocks, shear boundaries, reconnection, and other plasma processes via interactions with an opposing jet, ambient medium and external B-fields.

To address the relevance of our ring-laser jets to astrophysical jets, in Table 1 we highlight sample physical parameters of the d=800μm ring jet at 3ns and 3.5 mm from the target. As these hydrodynamic jets are kinetic-dominated (in contrast to magnetic-dominated jets driven by pulse power), they are most relevant to the study of YSO jets (Frank et al 2014). Table 1 shows that most dimensionless parameters (plasma $\beta$, flow $\sigma$, Mach number, Reynolds number, and magnetic Reynolds number) all lie in the same regime for both jets, suggesting the scalability of important properties of the ring-laser jets to the astrophysics regime (Ryutov et al 1999. 2000). It has been proposed that at least the stability of some YSO jets may be supported or enhanced by poloidal magnetic fields (Albertazzi et al 2014). Since a ring-laser jet is embedded with strong poloidal fields (Fig.4), it is an ideal platform to study the stabilizing effects of strong poloidal magnetic fields, with application to the stability of YSO jets. Since the density, temperature, flow speed and magnetic field of the ring-laser jet can be varied in a controllable manner by dialing the ring radius and laser parameters, a broad range of YSO jets with matching dimensionless parameters can be studied (Frank et al 2014). We emphasize that the ring-laser platform can be further expanded in many new directions. For example, by adding high-Z dopants to the CH target (cf. Fig.6), we can reduce the plasma $\beta(=p_{gas}/p_B)$ via enhanced radiative cooling and reach the low-$\beta$ regime, which may lead to new physics. We can add angular momentum to the jet by using tilted-tile target surfaces so that the blow off from each individual laser spot becomes left-right asymmetric. We can also replace the flat target with cone-shape target. Depending on the opening angle of the cone, the increased convergence of the on-axis flow should lead to even higher magnetic fields, density, temperature and flow speed, all of which can be controlled by dialing both the cone angle and ring radius.

Another important and unique property of these kinetic-dominated but strongly magnetized jets is their thermal conductivity (Spitzer 2006, Braginskii 1958). Electron transport in these jets becomes highly anisotropic due to the strong fields and small gyroradii (Table 1). As a result, electron heat conduction is suppressed orthogonal to **B** (Braginskii 1958), but remains Spitzer-like parallel to **B** (Spitzer 2006), leading to steep gradient for the electron temperature in the radial direction but less steep gradient in the axial direction (Fig.4b). Anisotropic thermal conduction is not yet incorporated in the FLASH code. But future FLASH predictions including anisotropic thermal conduction should be testable using our TS and XRFC data. The role of thermal conduction is an outstanding unsolved problem in many fields of astrophysics, including YSO jets. Our ring-laser platform can provide a new approach to study magnetized thermal conduction relevant to both laboratory and astrophysical plasmas.

This research was supported by DOE NNSA grant DE-NA0002721. FLASH simulations were performed on Extreme Science and Engineering Discovery Environment (XSEDE), which is supported by National Science Foundation grant number ACI-1548562, and the Argonne Leadership Computing Facility, which is a DOE Office of Science User Facility supported under Contract DE-AC02-06CH11357. Proton radiography was supported in part by DOE Grant No. DE-FG03-09NA29553, No.DE-SC0007168. EL and YL acknowledge partial support by LANL/LDRD during the writing of this paper.

**Table 1**. Parameters of the d=800 μm jet at 3.5 ns and 3.5 mm from laser target (left column) compared to those of young stellar object (YSO) jets (right column).

| d=800 μm OMEGA jet | YSO jet |
|---|---|
| Electron density $n_e \sim 1.5 \times 10^{20}$ cm$^{-3}$ | $\sim 10^2 - 10^5$ cm$^{-3}$ |
| Electron temperature $T_e \sim 1$ keV | $\sim 10^4$ – few x $10^6$ K |
| Ion temperature $T_i \sim 2.5$ keV | $\sim T_e$ |
| Ionization <Z> ~ 3.5 | low – 100% |
| Flow velocity v ~ $10^8$ cm/s | few x $10^7$ cm/s |
| Magnetic field B ~ $10^6$ Gauss | ~ 20 – 500 μG |
| Plasma Beta $\beta = 8\pi P_g/B^2 \sim 12$ | ~ 10 - $10^3$ |
| Flow Sigma $\sigma = B^2/4\pi\rho v^2 \sim 0.09$ | ~ $10^{-4}$ |
| Mach number $M=v/c_s \sim 3$ | few - 10 |
| Alfven Mach number $M_A = v/v_A \sim 3.3$ | ~ $10^2$ |
| Reynolds number $R_e \sim 260$ | ~ 10 - $10^3$ |
| Magnetic Reynolds number $R_{eM} \sim 10^4$ | ~ few x $10^2$ |
| Peclet number $P_e = 1.5kn_evd/\kappa_T \sim 0.3$ | unknown |
| Electron skin depth $c/\omega_e \sim 0.4$ μm | |
| Ion skin depth $c/\omega_I \sim 24$ μm | |
| Debye length $v_e/\omega_e \sim 0.01$ μm | |
| Electron gyroradius $v_e/\omega_{Be} \sim 0.6$ μm | |
| Ion gyroradius $v_i/\omega_{Bi} \sim 20$ μm | |
| Coulomb scattering mean free path $\lambda_{ei} \sim 20$ μm | |

**Figure Captions**

Fig.1 Setup of the OMEGA magnetized jet experiment. Laser parameters and top-view of 20-beam pattern (colors denote different incident angles) are illustrated in the lower figure. The CH target is located 2.5 mm below target chamber center (TCC). $D^3He$ or OMEGA-EP proton sources are located ~1cm to the left of TCC and proton images are recorded at 16.5 – 17 cm to the right of TCC. A $2\omega_o$ TS probe beam measured plasma parameters at TCC with ~50 μm spatial resolution. Not shown is the XRFC, which images the jet at ~38$^o$ from the jet axis.

Fig.2 Comparison of P-rad images for jets launched by four different ring-laser radii d: (a) d=0, t=3.6ns; (b) d=400μm, t=2.6ns; (c) d=800μm, t=3.6ns; (d) d=1200μm, t=4.3ns. All images are those of 14.7-15 MeV protons. The circular cutoff at the bottom corresponds to the edge of the target.

Fig.3 Cross-section of integrated **B**-field profile (c) of the d=800 mm jet obtained from the direct inversion of the proton density profile (b), corresponding to the blue line-out in the P-rad image (a).

Fig.4 (a) Comparison of 3D FLASH simulations for |**B**| profiles at 3ns for four different ring-jet radii d. (b) Sample field lines at t=1.6ns, 2.8ns, and 3.6ns of the d=800μm jet. Color scales denote field amplitudes in kG. (c) Comparison of 3D FLASH-predicted P-rad images (left column) and observed $D^3He$ P-rad data (right column) for the d=800μm jet at the same times as Fig.4(b).

Fig.5 Top row: TS electron plasma wave (EPW) and ion acoustic wave (IAW) spectra vs. time for the d=400μm jet (combination of 3 shots). Rows 2 to 5: (left column) Time histories for density, flow velocity, ion and electron temperatures at TCC for jets of different ring radii d

derived from TS spectra. (right column) Time histories for density, flow velocity, ion and electron temperatures at TCC based on 3D FLASH simulations.

Fig.6 Time-lapse XRFC images (500ps exposure) of d= 800 μm jets for (a) 2% Fe-doped CH target and (b) pure CH target. In (a) the frames are taken at 2ns, 3ns, 4ns and 5ns. In (b) the frames are taken at 2ns, 2.5ns, 3ns, and 4 ns. The blue marker in each picture denotes a length of 1.6 mm. It is clear that the Fe-doped jet is narrower than the pure CH jet.

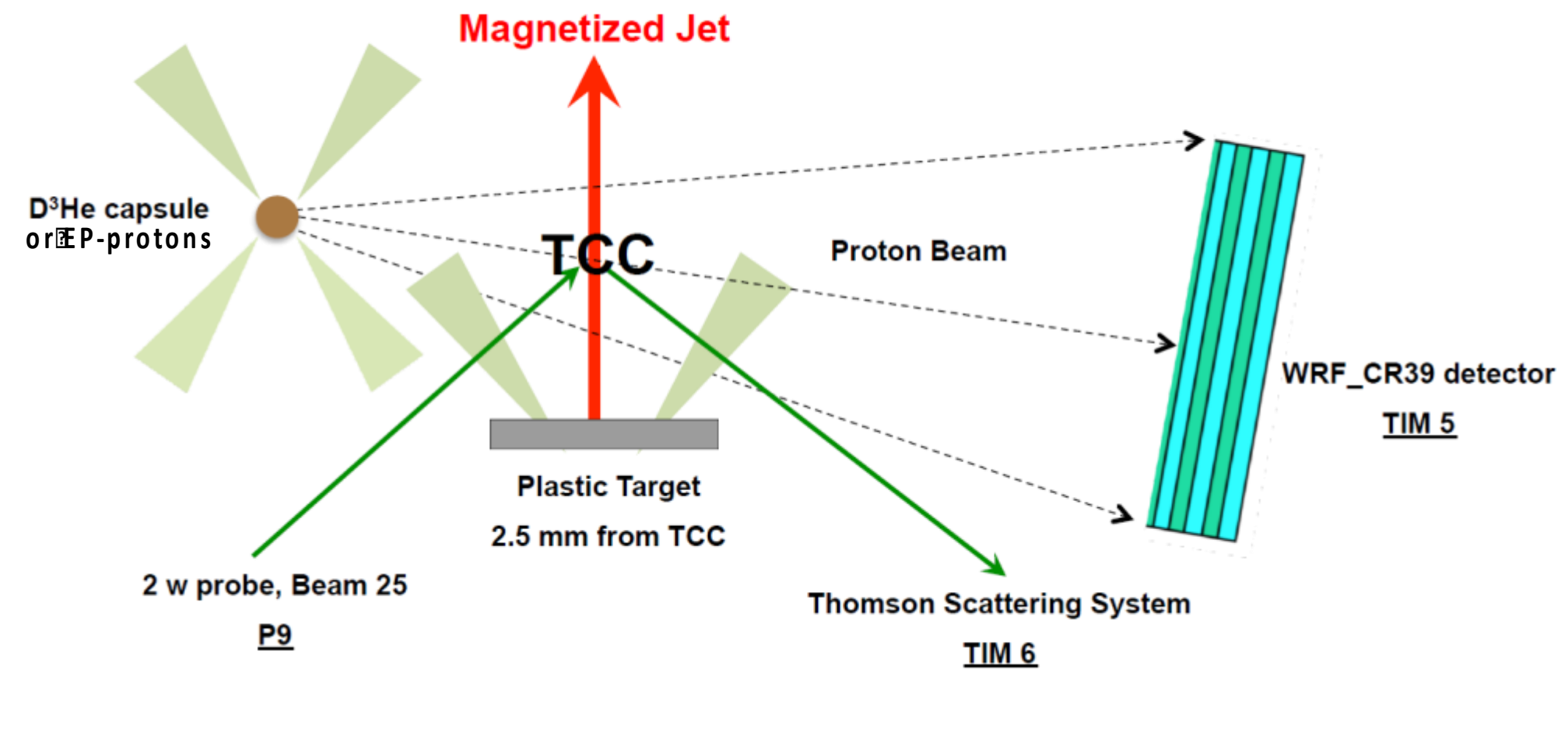
Side-on view:
Magnetized Jet
D³He capsule
or EP-protons
TCC
Proton Beam
WRF_CR39 detector
TIM 5
Plastic Target
2.5 mm from TCC
2 w probe, Beam 25
P9
Thomson Scattering System
TIM 6

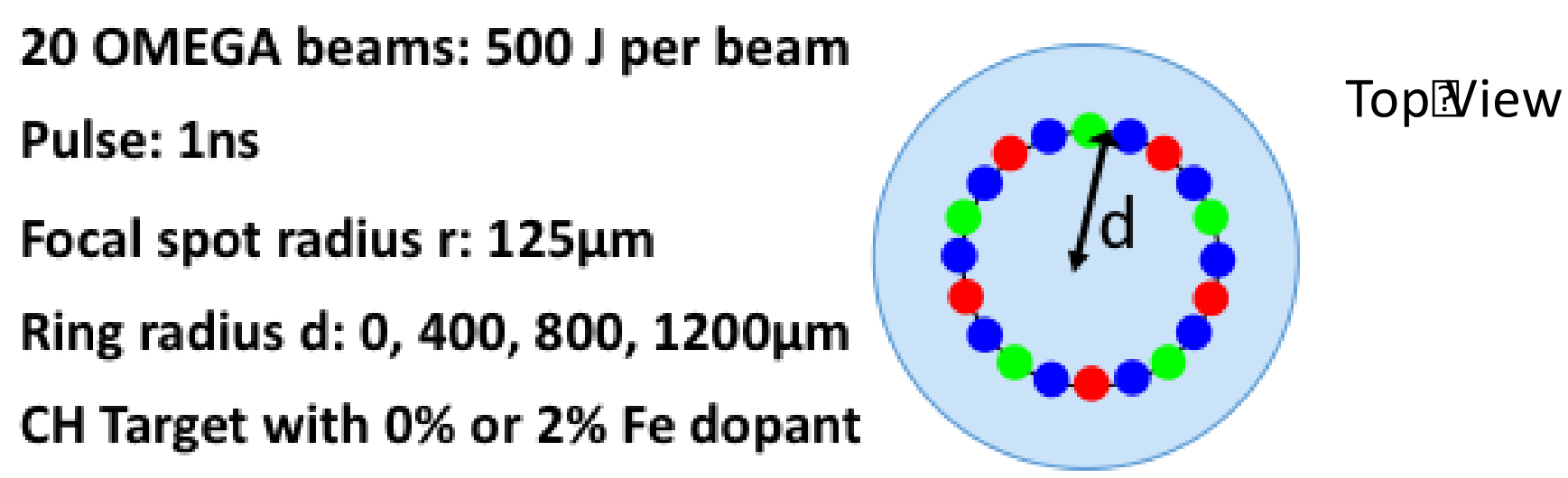
20 OMEGA beams: 500 J per beam
Pulse: 1ns
Focal spot radius r: 125µm
Ring radius d: 0, 400, 800, 1200µm
CH Target with 0% or 2% Fe dopant
d
Top View

Fig.1

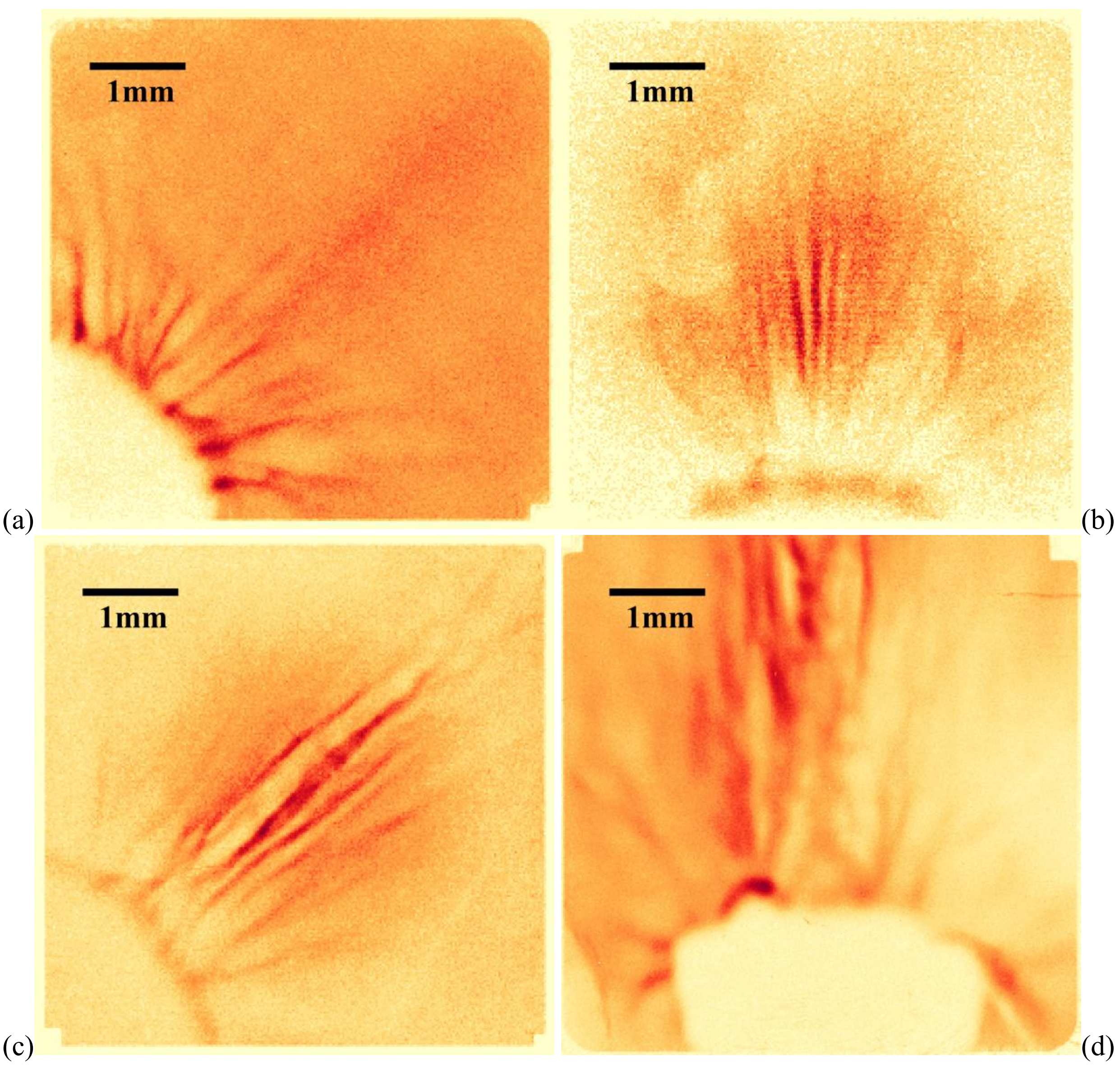

Fig.2

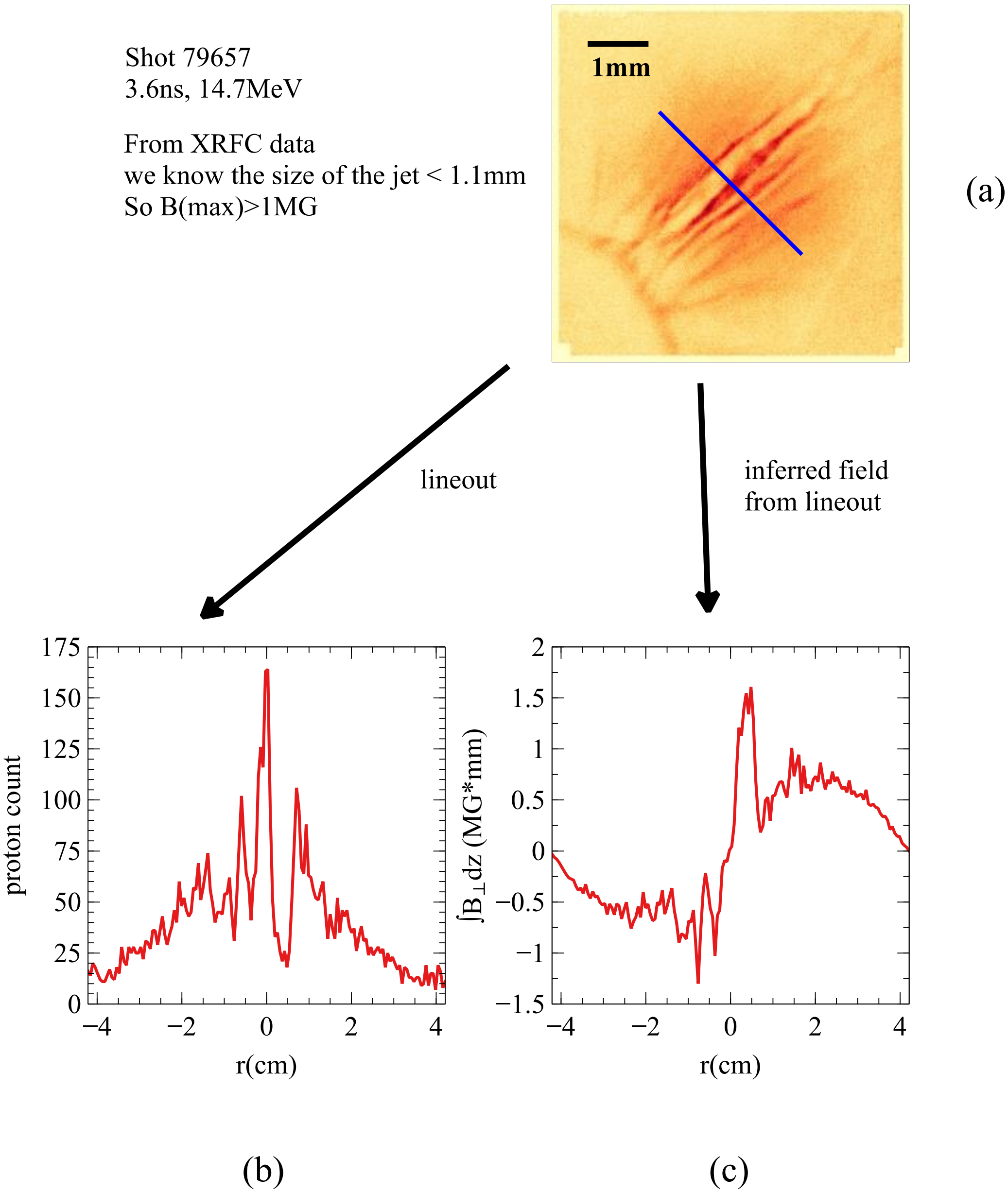
Shot 79657
3.6ns, 14.7MeV
From XRFC data
we know the size of the jet < 1.1mm
So B(max)>1MG
1mm
(a)
lineout
inferred field
from lineout
proton count
r(cm)
∫B⊥dz (MG*mm)
r(cm)
(b)
(c)

Fig.3

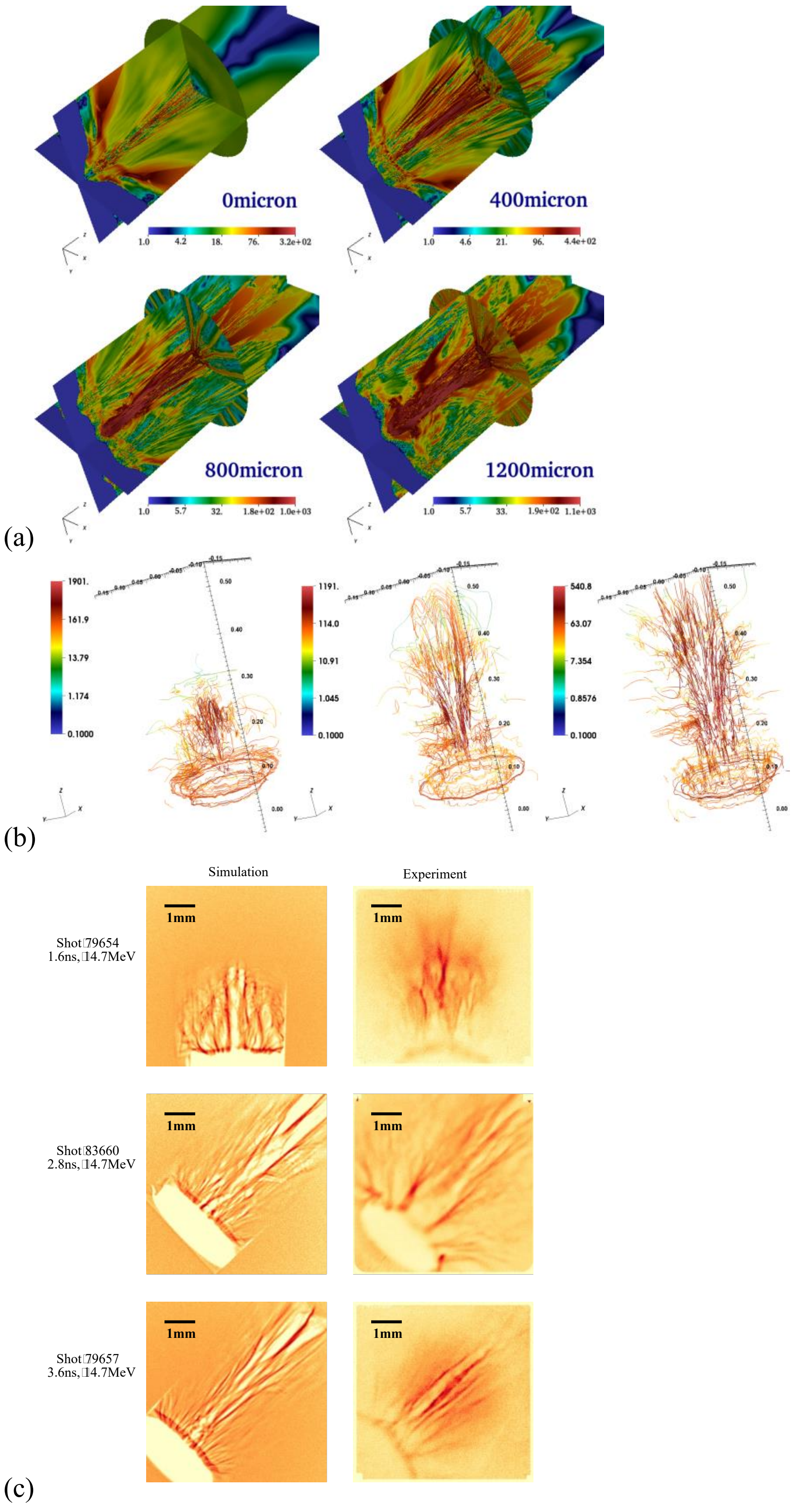
0micron
400micron
800micron
1200micron
(a)
(b)
Simulation
Experiment
1mm
Shot 79654
1.6ns, 14.7MeV
Shot 83660
2.8ns, 14.7MeV
Shot 79657
3.6ns, 14.7MeV
(c)

Fig.4

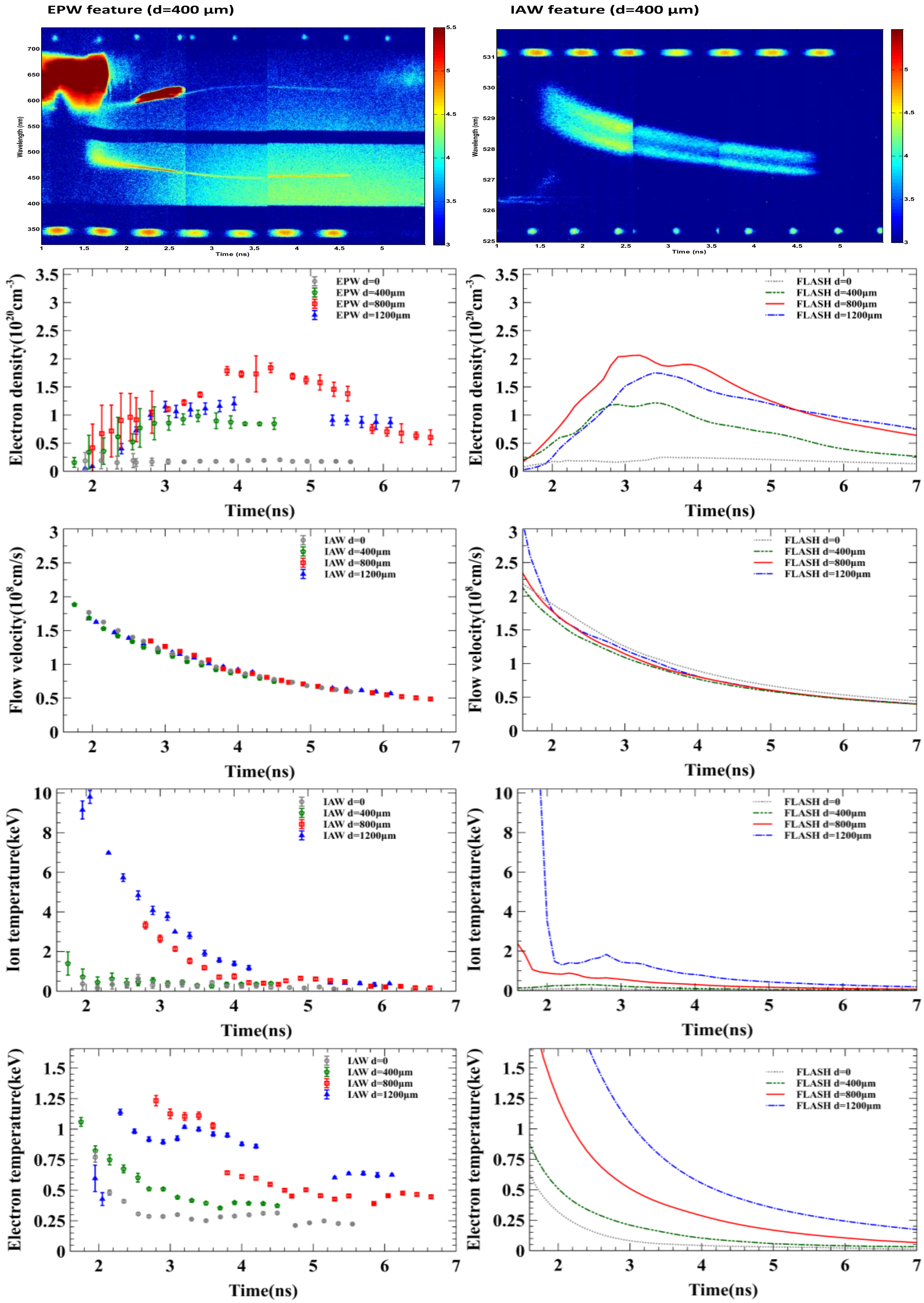

EPW feature (d=400 μm)
IAW feature (d=400 μm)
Wavelength (nm)
Time (ns)
EPW d=0
EPW d=400μm
EPW d=800μm
EPW d=1200μm
FLASH d=0
FLASH d=400μm
FLASH d=800μm
FLASH d=1200μm
IAW d=0
IAW d=400μm
IAW d=800μm
IAW d=1200μm
Electron density($10^{20}$cm$^{-3}$)
Flow velocity($10^{8}$cm/s)
Ion temperature(keV)
Electron temperature(keV)
Time(ns)

Fig.5

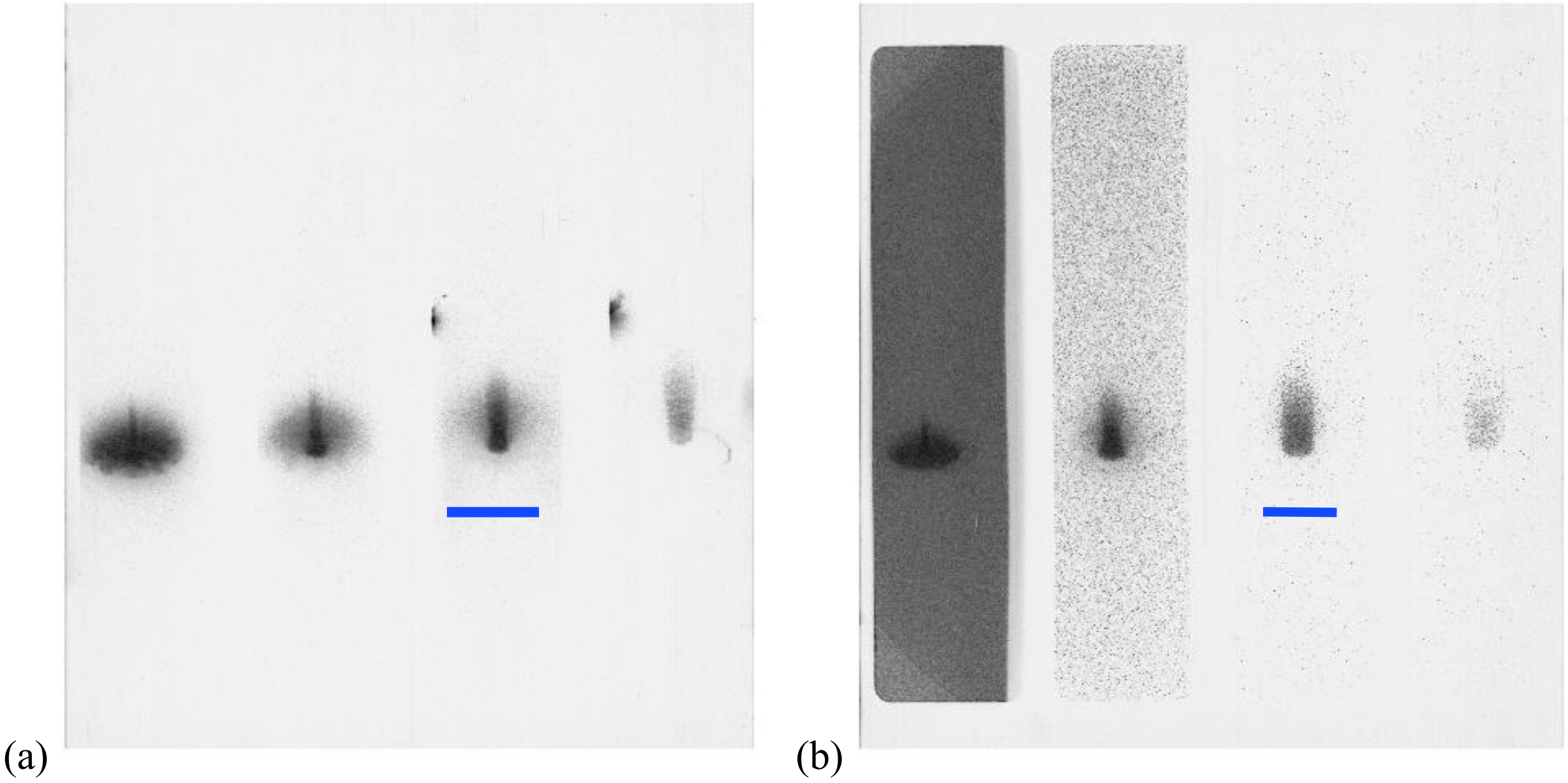

(a) (b)

Fig.6